# Consistent multiple-relaxation-time lattice Boltzmann method for the volume-averaged Navier-Stokes equations

Yang Liu[1], Xuan Zhang[2], Jingchun Min[1, *], Xiaomin Wu[3, *]

[1] Department of Engineering Mechanics, Tsinghua University, Beijing 100084, China
[2] Department of Energy and Power Engineering, Beijing Institute of Technology, Beijing 100081, China
[3] Department of Energy and Power Engineering, Tsinghua University, Beijing 100084, China
*Corresponding author, Email: minjc@mail.tsinghua.edu.cn, wuxiaomin@mail.tsinghua.edu.cn

**Abstract:** Recently, we notice that a pressure-based lattice Boltzmann (LB) method was established to recover the volume-averaged Navier-Stokes equations (VANSE) [1], which serve as the cornerstone of various fluid-solid multiphase models. It decouples the pressure from density and exhibits excellent numerical performance, however, the widely adopted density-based LB scheme still suffers from significant spurious velocities and inconsistency with VANSE. To remedy this issue, a multiple-relaxation-time LB method is devised in this work, which incorporates a provisional equation of state in an adjusted density equilibrium distribution to decouple the void fraction from density. The Galilean invariance of the recovered VANSE is guaranteed by introducing a penalty source term in moment space, effectively eliminating unwanted numerical errors. Through the Chapman-Enskog analysis and detailed numerical validations, this novel method is proved to be capable of recovering VANSE with second-order accuracy consistently, and well-suited for handling void fraction fields with large gradients and spatiotemporal distributions.

## 1. Introduction

The volume-averaged Navier-Stokes equations (VANSE), bridging the microscopic interactions with the macroscale flow behavior, are crux for the coarse-grained simulations of fluid-solid multiphase systems [2]. These equations can be derived from the standard Navier-Stokes equations via the local averaging technique, which introduces the void fraction to describe the proportions of fluid and solid phases per unit volume disregarding of their detailed interfaces [3]. By upscaling the microstructural effects into theoretical or empirical drag and collision laws from the unresolved mesoscopic perspective, several multiphase approaches rooted in VANSE have garnered attention in recent years for their computational efficiency and satisfactory accuracy in particulate tracking [4]. For dispersed multiphase flows, VANSE are integral to two prevalent frameworks: the two-fluid model (TFM) within the Euler-Euler framework, and the discrete-element method (DEM) or the particle-in-cell model within the Euler-Lagrange framework [4-6]. These methods provide feasible numerical tools for simulating complex flow phenomenon involving a large number of particles or bubbles, e.g., fluidized bed [7], blood cell flow [8], and microgel suspension [9]. Furthermore, VANSE also contribute to underpin the well-known Darcy-Brinkman-Stokes (DBS) approach within the micro-continuum framework, significantly advancing the multiscale modeling of flow behavior at the representative-elementary-volume (REV) scale [10]. When coupled with a range of multi-physical processes (e.g., interface tracking methods, species transport equations, poroelasticity methodologies and crystal nucleation theories), the DBS framework has provided promising paths for addressing various challenges of porous flow systems, including mineral dissolution [11], coke combustion [12], matrix deformation [13] and capillary/viscous fracturing [14].

As a mesoscopic numerical tool that has made outstanding progress in modeling multiphase flow, the lattice Boltzmann method (LBM) recently has also been extended to solve VANSE systems, due to its notable advantages in handling complex interfaces, ease of implementation and flexibility in parallel computing [15]. From a bottom-up approach, the so-called gray-LB model was proposed to simulate fluid behavior in the presence of sub-resolution solid components, which modified the LB propagation step by introducing a partial bounce-back fraction (corresponding to the void fraction in VANSE) to mimic local fluid-solid interactions [16-18]. However, the numerically obtained fluid-solid effects were viscosity-dependent and non-physical, the gray-LB model was therefore incompatible with VANSE as it failed to accurately reproduce macroscale behaviors [19].

On the other hand, a series of numerical studies on building VANSE-LB schemes following the up-bottom concept have made progress. Notably, Guo et al. [20] pioneered a REV-scale LB model recovering the simplified version of VANSE for porous flows. Although this model was primarily effective for cases with uniform void fraction fields, it had been widely extended in various applicable scenarios, e.g., immiscible multiphase seepage [21], fluid flow through moving porous media [22], and frost growth on cold surfaces [23]. To the best of our knowledge, the prototype of the density-based LB scheme for the full form of VANSE was first proposed by Wang et al. [24] and later theoretically validated by Zhang et al [25], which incorporated a correction force to modify the pressure gradient term. However, according to the detailed numerical investigation by Blais et al. [26], this correction force could induce significant non-physical pressure jumps and spurious velocities in heterogeneous void fraction fields. Despite these shortcomings, this scheme has been preliminarily used to develop TFM-LBMs for systems with small void fraction gradients [27, 28]. To this end, Blais et al. [26] also constructed a source term for the correct deviatoric stress tensor, and then introduced a novel collision operator to improve the stability of pressure calculations for very large void fraction gradients, but the solutions from this treatment diverge when the void fraction falls below 0.5. Alternatively, Höcker et al. [29] eliminated the non-physical pressure oscillation by manipulating the LB streaming step, thereby avoiding the need for force correction, and this approach was successfully implanted into DEM-LBM to model reactive particle-fluid flows [30]. Unfortunately, the recovered continuity equation and viscous term remained inconsistent with those in VANSE. Subsequently, Bukreev et al. [31] redefined the zeroth moment of population distribution (or the density equilibrium distribution) using the locally-integral values of void fraction, thus recovering VANSE with second-order convergence, while this algorithm still encountered spurious velocities at discontinuous distributions. Most recently, Fu et al. [1] established a pressure-based LB scheme decoupling pressure from density, which effectively reduced unwanted spurious velocities and yielded second-order convergence in VANSE recovery. Although this scheme was validated to perform well for spatiotemporal void fraction fields, our simulations elucidate that the use of single-relaxation-time (SRT) collision operator limits its applicability for a larger viscosity range, which will be given in the following. Moreover, considering that the density evolution models are mainstream in current LB community, developing a consistent density-based VANSE-LB scheme is still highly necessary.

In this work, a novel multiple-relaxation-time lattice Boltzmann method for VANSE (MRTLB-VANSE) is devised within the density-based framework. Specifically, a provisional equation of state is introduced to decouple the void fraction from density by modifying the density equilibrium distribution. Although this scheme still employs a correction force to address the inconsistency of the pressure gradient term, it avoids the need to discretize the void fraction gradient in force calculations. Additionally, inspired by the works of Li et al. [32, 33] and confirmed by Chapman-Enskog analysis, the MRT collision operator allows for the construction of a penalty source term in moment space, effectively eliminating the numerical errors of viscous stress tensor in the recovered macroscopic momentum. Accordingly, this paper is structured as follows: Section 2 introduces the general form of VANSE, discusses the numerical errors and failure reasons in previous VANSE-LB models, and devises the specific framework of MRTLB-VANSE; Section 3 conducts several numerical validations, including benchmarks against analytical solutions, spurious velocity tests, and convergence verification of VANSE for spatial and temporal varying void fraction fields; Section 4 concludes this paper culminating with future prospects; At the end, Appendix A details the Chapman-Enskog analysis, confirming the devised scheme's consistency with VANSE; Appendix B additionally presents the second-order equilibrium distribution and the penalty source term within the MRT-VANSE framework; Appendix C gives three specific expressions of body force used for convergence verification.

## 2. Methodology
### 2.1. Volume-averaged Navier-Stokes equations

To date, several versions of VANSE have been proposed based on different concepts of volume-averaging and approaches to solving fluid stress tensor, as thoroughly reviewed by Zhou et al. [2] In this paper, without loss of generality, the formulation of so-called model A (set II) is adopted, since it could be transformed into model B (set I) via some special treatments [34, 35]. The continuity and momentum equations of model A can be expressed as

$$\partial_t(\phi\rho) + \nabla \cdot (\phi\rho \mathbf{u}) = 0,$$
$$\partial_t(\phi\rho\mathbf{u}) + \nabla \cdot (\phi\rho\mathbf{u}\mathbf{u}) = -\phi\nabla p + \nabla \cdot (\phi\boldsymbol{\tau}) + \mathbf{F}, \quad (1)$$

where $\phi$, $\rho$, $p$, $\mathbf{u}$ are the void fraction (porosity), density, pressure, and velocity of fluid; $\mathbf{F}$ is the body force term containing drag laws for fluid-solid coupling and other external forces; $\boldsymbol{\tau}$ is the viscous stress tensor for Newtonian fluids

$$\boldsymbol{\tau} = \rho\upsilon\left((\nabla\mathbf{u}) + (\nabla\mathbf{u})^{\mathrm{T}} - \frac{2}{3}(\nabla\cdot\mathbf{u})\mathbf{I}\right), \quad (2)$$

where $\upsilon$ is kinematic viscosity. It is noted that the last term of right hand cannot be omitted here since the divergence of velocity for VANSE is nonzero in a nonuniform void fraction field.

### 2.2. Numerical error of BGK-LB method for VANSE

The LB scheme with the Bhatnagar-Gross-Krook (BGK) collision operator for recovering VANSE was theoretically given by Zhang et al.[25], which evolutes the local information of fluid mass via redefining the density distribution function in the discrete LB equation, given by

$$f_i(\mathbf{x}+\mathbf{e}_i\delta_t, t+\delta_t) = f_i(\mathbf{x},t) - \tau_\upsilon\left(f_i(\mathbf{x},t) - f_i^{eq}(\mathbf{x},t)\right) + \delta_t\left(1 - \frac{\tau_\upsilon}{2}\right)F_i(\mathbf{x},t), \quad (3)$$

where $f_i(\mathbf{x},t)$ is the population distribution at the position $\mathbf{x}$ and time $t$ in the $i$-th discrete direction, $f_i(\mathbf{x}+\mathbf{e}_i\delta_t, t+\delta_t)$ is the post-collision distribution, and the equilibrium distribution $f_i^{eq}$ is defined as

$$f_i^{eq}(\mathbf{x},t) = \omega_i\phi\rho\left(1 + \frac{\mathbf{e}_i\cdot\mathbf{u}}{c_s^2} + \frac{(\mathbf{e}_i\cdot\mathbf{u})^2}{2c_s^4} - \frac{\mathbf{u}\cdot\mathbf{u}}{2c_s^2}\right), \quad (4)$$

where the rescaling of density $\rho$ into $\phi\rho$ allows the LB equation to be correlated from NSE to VANSE, and the lattice sound speed is given by $c_s = c/\sqrt{3}$, bridging the relaxation factor $\tau_\upsilon$ with the shear viscosity $\upsilon$, i.e., $\tau_\upsilon^{-1} = \upsilon/c_s^2\delta_t + 0.5$. The D2Q9 lattice stencil with a constant lattice speed $c = \delta_x/\delta_t = 1$ is employed in this paper, and it can straightforwardly be extended to 3D and other lattices. The discrete lattice velocity $\mathbf{e}_i$ and the corresponding weight coefficients $\omega_i$ are provided as

$$\mathbf{e}_i = \begin{cases} (0,0), & \omega_{i=0} = 4/9 \\ (\cos[(i-1)\pi/2], \sin[(i-1)\pi/2])c, & \omega_{i=1\sim4} = 1/9 \\ (\cos[(2i-9)\pi/4], \sin[(2i-9)\pi/4])\sqrt{2}c, & \omega_{i=5\sim8} = 1/36 \end{cases}. \quad (5)$$

Furthermore, the force term $F_i$ that takes account of an additional term $\mathbf{F}^P = \rho c_s^2 \nabla \phi$ responsible for correcting the pressure gradient in VANSE, as well as the external body force $\mathbf{F}^B$, is given with the scheme of Guo et al. [36]

$$F_i(\mathbf{x},t) = \omega_i \left( \frac{\mathbf{e}_i - \mathbf{u}}{c_s^2} + \frac{\mathbf{e}_i \cdot \mathbf{u}}{c_s^4} \mathbf{e}_i \right) \cdot (\mathbf{F}^P + \mathbf{F}^B). \tag{6}$$

Then, the macroscopic fluid density, velocity, and pressure can be obtained via the moments of the distribution function

$$\rho = \sum_i f_i / \phi, \quad \mathbf{u} = \left( \sum_i \mathbf{e}_i f_i + \frac{\delta_t}{2} (\mathbf{F}^P + \mathbf{F}^B) \right) \Big/ \sum_i f_i, \quad p = \rho c_s^2. \tag{7}$$

Accordingly, through the third-order Chapman-Enskog analysis of Eq. (3), the following macroscopic equations can be recovered when the external body force $\mathbf{F}^B$ is ignored [37]

$$\partial_t (\phi \rho) + \nabla \cdot (\phi \rho \mathbf{u}) = 0, \tag{8}$$

$$\partial_t (\phi \rho \mathbf{u}) + \nabla \cdot (\phi \rho \mathbf{u}\mathbf{u}) = -\nabla (\phi p) + \mathbf{F}^P + \upsilon \nabla \cdot (\phi \tau) + \frac{\delta_t^2}{12} \nabla \cdot \nabla \mathbf{F}^P. \tag{9}$$

The last term on the right side of Eq. (9) is the high-order error that inevitably occurs when the pressure correction is applied in force term. Moreover, it should be noted that in discrete space, the value of "$-\nabla(\phi p) + \mathbf{F}^P$" is not exactly equal to the correct pressure term "$-\phi \nabla p$" as the continuous derivative. In this paper, the isotropic finite difference stencil is used to evaluate the gradient operator with [38]

$$\partial_\alpha \psi(\mathbf{x}) = \frac{1}{c_s^2 \delta_t} \sum_i \omega_i \psi(\mathbf{x} + \mathbf{e}_i \delta_t) \mathbf{e}_{i\alpha}, \tag{10}$$

where $\psi$ denotes an arbitrary variable that requires discretization. Through Taylor expansion for this difference stencil in D2Q9 space, the discrete form of the correction force is deduced

$$\overline{\mathbf{F}}^P = \mathbf{F}^P + \frac{\delta_t^2}{6} \nabla (\nabla \cdot \mathbf{F}^P) + \cdots. \tag{11}$$

Assuming that the influence of weak compressibility in LB scheme is negligible, the total numerical error below the fourth-order combining Eq. (9) and Eq. (11) is given by

$$\mathbf{F}^P_{error} = \frac{\delta_t^2}{12} \nabla \cdot \nabla \mathbf{F}^P + \frac{\delta_t^2}{6} \nabla (\nabla \cdot \mathbf{F}^P) = \frac{\rho}{12} \nabla (\nabla^2 \phi). \tag{12}$$

Note that this error cannot be ignored when $\phi$ is distributed with large gradients, as it could lead to non-physical acceleration towards the flow field, resulting in spurious velocities that exceed the Mach number limit of LBM. This is the primary issue of numerical instability and inconsistency of the previous VANSE-LB scheme for spatiotemporal void fraction field.

### 2.3. Consistent MRT-LB method for VANSE

The consistent LB scheme presented in this section still uses the forcing term for pressure correction towards VANSE. Fundamentally speaking, the use of $\mathbf{F}^P = \rho c_s^2 \nabla \phi$ in Eq. (6) originates from the definition of $\phi \rho$ in Eq. (4), which recovers the equation of state $p^{EOS} = \phi \rho c_s^2$ within the inconsistent pressure gradient term of Eq. (9). To address this issue, a provisional equation of state $p_0^{EOS} = \kappa \rho c_s^2$ is adopted to replace $\nabla p^{EOS}$ with $\nabla p_0^{EOS}$ in Eq. (9). Meanwhile, the pressure correction force is readjusted into

$$\mathbf{F}^p = \nabla(\kappa \rho c_s^2) - \phi \nabla(\rho c_s^2) = (\kappa - \phi) c_s^2 \nabla \rho, \tag{13}$$

where $\phi$ and $\rho$ are decoupled in gradient operator, $0 \leq \kappa \leq 1$ is a constant suggested to choose close to the minimum void fraction of the whole domain. Obviously, this force term is independent of $\nabla \phi$, and its value converges to zero after a certain calculation step for incompressible flows with constant density. Even for multiphase flow, the dissipative interfaces in LBM causes density to typically vary among several lattice nodes, instead of intermittent distribution. Hence, this setup can effectively avoid non-physical oscillations at the discontinuities of void fraction field.

In order to yield the provisional equation of state mentioned above, the equilibrium density function built from the 3rd-order Hermite expansion of Maxwellian distribution is adopted here [32, 39, 40]

$$f_i^{eq}(\mathbf{x},t) = \begin{cases} \phi\rho - \sum_{i \neq 0}\omega_i \kappa\rho - \omega_0 \phi\rho \dfrac{\mathbf{u}\cdot\mathbf{u}}{2c_s^2}, & i = 0 \\ \omega_i\left(\kappa\rho + \phi\rho\left(\dfrac{\mathbf{e}_i\cdot\mathbf{u}}{c_s^2}\left(1 + \dfrac{1}{2}(X-1)(\dfrac{\mathbf{e}_i^2}{c_s^2} - 4)\right) + \dfrac{(\mathbf{e}_i\cdot\mathbf{u})^2}{2c_s^4} - \dfrac{\mathbf{u}\cdot\mathbf{u}}{2c_s^2}\right)\right), & i = 1 \sim 8 \end{cases} \tag{14}$$

where $X = P_0^{EOS}/P^{EOS} = \kappa/\phi$, and the 0th- to 3rd-order velocity moments of this equilibrium distribution can be calculated as follows

$$\begin{aligned}
&\sum_i f_i^{eq} = \phi\rho, \\
&\sum_i f_i^{eq} e_{i\alpha} = \phi\rho u_\alpha, \\
&\sum_i f_i^{eq} e_{i\alpha} e_{i\beta} = \phi\rho u_\alpha u_\beta + \kappa\rho c_s^2 \delta_{\alpha\beta}, \\
&\sum_i f_i^{eq} e_{i\alpha} e_{i\beta} e_{i\gamma} = \begin{cases} \phi\rho c_s^2 (u_\gamma \delta_{\alpha\beta} + u_\beta \delta_{\alpha\gamma} + u_\alpha \delta_{\beta\gamma}), & \text{if } i = j = k, \\ \kappa\rho c_s^2 (u_\gamma \delta_{\alpha\beta} + u_\beta \delta_{\alpha\gamma} + u_\alpha \delta_{\beta\gamma}), & \text{otherswise}, \end{cases}
\end{aligned} \tag{15}$$

where $\delta$ is the Kronecker delta with two indices. Evidently, the 0th and 1st-order moments are identic with the VANSE-LB scheme given by Zhang et al. [25], so the macro parameters are still calculated by Eq. (7), meanwhile, the pressure term in the 2nd-order moment is successfully modified. However, in the 3rd-order moment, the introduction of the provisional equation of state produces bias that brings numerical errors in the recovered viscous stress tensor, thereby making the Galilean invariance of the momentum equations violated [41, 42]. Consequently, motived by the pioneer work of Li et al. [32], we employ the MRT collision operator in this paper to eliminate the unwanted error via adding a source term in moment space. This strategy has also been utilized to treat such issue in color-gradient and free-energy multiphase LB models [33, 40, 43, 44]. Besides, the use of MRT scheme for VANSE has more robust numerical performance compared with SRT, which will be demonstrated in the following.

Multiplying Eq. (3) with the transformation matrix $\mathbf{M}$ cf. Appendix A, the evolution equation of MRTLB-VANSE is expressed as

$$\mathbf{m}^+ = \mathbf{m} - \mathbf{\Gamma}(\mathbf{m} - \mathbf{m}^{eq}) + \delta_t \left(\mathbf{I} - \frac{\mathbf{\Gamma}}{2}\right)(\mathbf{S} + \mathbf{C}), \tag{16}$$

where the population distribution in moment space is mapped by $m_i = \sum_j \mathbf{M}_{ij} f_j$, written as

$$\mathbf{m} = \mathbf{M}\mathbf{f} = (m_0, m_1, \cdots, m_8)^{\mathrm{T}} = (\phi\rho, e, \varepsilon, j_x, q_x, j_y, q_y, p_{xx}, p_{xy})^{\mathrm{T}}, \tag{17}$$

in which $e$ is the energy mode, $\varepsilon$ is related to the energy square, $(q_x, q_y)$ are the energy flux, $(j_x, j_y) = (\phi\rho u_x - \delta_t F_x/2, \phi\rho u_y - \delta_t F_y/2)$ are the components of momentum, and $(p_{xx}, p_{xy})$ correspond to the diagonal and off-diagonal components of the stress tensors [45]. Based on Eq. (14), the equilibrium moment $m_i^{eq} = \sum_{j,k} \mathbf{M}_{ij} f_j^{eq}$ can be explicitly given by

$$\begin{aligned}\mathbf{m}^{eq} &= \mathbf{M}\mathbf{f}^{eq} = (m_0^{eq}, m_1^{eq}, \cdots, m_8^{eq})^{\mathrm{T}} = \\ &\phi\rho\left(1, -4 + 3\mathbf{u}^2 + 2X, 4 - 3\mathbf{u}^2 - 3X, u_x, (X-2)u_x, u_y, (X-2)u_y, u_x^2 - u_y^2, u_x u_y\right)^{\mathrm{T}},\end{aligned} \tag{18}$$

Similarly, the force moment $S_i = \sum_{j,k} \mathbf{M}_{ij} F_j$ can be derived from Eq. (6), expressed as

$$\begin{aligned}\mathbf{S} &= \mathbf{M}\mathbf{F} = (S_0, S_1, \cdots, S_8)^{\mathrm{T}} = \\ &\left(0, 6\mathbf{u}\cdot\mathbf{F}, -6\mathbf{u}\cdot\mathbf{F}, F_x, -F_x, F_y, -F_y, 2(u_x F_x - u_y F_y), u_x F_y + u_y F_x\right)^{\mathrm{T}}.\end{aligned} \tag{19}$$

Moreover, $\mathbf{m}^+$ is the moment distribution required to project back discrete space for the streaming step, i.e., $f_i(\mathbf{x} + \mathbf{e}_i\delta_t, t + \delta_t) = f_i^+(\mathbf{x}, t)$, therein $\mathbf{f}^+ = \mathbf{M}^{-1}\mathbf{m}^+$ and $\mathbf{M}^{-1}$ is the inverse matrix of $\mathbf{M}$, cf. Appendix A. $\mathbf{I}$ is the unit matrix; $\mathbf{\Gamma} = \mathrm{diag}(\tau_c, \tau_e, \tau_\varepsilon, \tau_c, \tau_q, \tau_c, \tau_q, \tau_\upsilon, \tau_\upsilon)$ is the non-negative relaxation matrix, therein $\tau_c = 1$ is chosen for density and momentum conservation, $\tau_\varepsilon = 1.2, \tau_q = 1.4$ are chosen for numerical stability requirements, and $\tau_e$ is related to the bulk viscosity $\upsilon_e$, i.e., $\tau_e = (\upsilon_e/c_s^2\delta_t + 0.5)^{-1}$; $\mathbf{C} = (0, C_a, 0, 0, 0, 0, 0, C_b, C_c)^{\mathrm{T}}$ is the penalty source term for viscous stress correction composed by

$$\begin{aligned}C_a &= -\kappa\left(\partial_x(\rho u_x) + \partial_y(\rho u_y)\right) + 2(\kappa - \phi)\rho(\partial_x u_x + \partial_y u_y) \\ &\quad + (3 - 2X)\left(\partial_x(\phi\rho u_x) + \partial_y(\phi\rho u_y)\right) - 2X\rho\partial_t\phi,\end{aligned} \tag{20}$$

$$\begin{aligned}C_b &= -\kappa\left(\partial_x(\rho u_x) - \partial_y(\rho u_y)\right) + \frac{2}{3}(\kappa - \phi)\rho(\partial_x u_x - \partial_y u_y) \\ &\quad + \left(\partial_x(\phi\rho u_x) - \partial_y(\phi\rho u_y)\right),\end{aligned}$$

$$C_c = \frac{1}{3}(\kappa - \phi)\rho(\partial_x u_y + \partial_y u_x),$$

where the spatial derivative $\nabla = (\partial_x, \partial_y)$ is calculated by Eq. (10), the time derivative is evaluated by the first-order Euler scheme, i.e., $\partial_t\phi = \left(\phi(\mathbf{x}, t + \delta_t) - \phi(\mathbf{x}, t)\right)/\delta_t$.

Based on the above, this designed MRT-LB scheme can exactly recover the consistent VANSE, which is confirmed by Chapman-Enskog analysis, cf. Appendix A. Notably, if we use $\mathbf{F}^p = (\kappa - 1)c_s^2\nabla\rho$ for pressure correction instead of Eq. (13) and change the term "$\kappa - \phi$" to "$\kappa-1$" in the source term of Eq. (20), the model B (set I) version of VANSE can be recovered with correct pressure gradient term and viscous stress tensor. Additionally, in framework of the MRT scheme mentioned above, we also provide the scheme of widely-adopted 2nd-order equilibrium distribution and its penalty source term, cf. Appendix B. They make the same effects in Chapman-Enskog analysis as derived in Appendix A to recover VANSE, without losing the consistency and accuracy. However, it is observed that the gradient calculation in Eq. (20) is reduced compared with Eq. (B3) in Appendix B, which is more advantageous for computational efficiency and numerical convergence. Thus, we still use the MRTLB-VANSE scheme with the 3rd-order equilibrium distribution henceforth for further validation.

### 3. Numerical validation

In this section, the physical correctness of the novel MRTLB-VANSE is validated against analytical solutions of several typical benchmark scenarios, and its convergence accuracy has been tested by the method of manufactured solutions (MMS). The computational domains in all test cases are discretized by the D2Q9 lattice space with periodic boundaries unless otherwise specified, and for default settings, the fluid density $\rho$ is initialized as 1.0, the kinematic viscosity $\upsilon$ is 0.1, and the constant $\kappa$ is chosen as 0.5.

### 3.1 Uniform porous flow

Accurately describing the flow behavior within a homogeneous substrate with uniform porosity field is a basic capability of VANSE. Neglecting the nonlinear Forchheimer force that characterizes inertial effects, and incorporating the external force $G_x$ and the linear Darcy resistance $\phi \upsilon u k_0^{-1}$ induced by fluid-solid interactions within the body force term in Eq. (6), i.e., $F_x^B = \phi(G_x - \phi \upsilon u k_0^{-1})$, VANSE can be reduced to the Darcy-Brinkman equation governing REV-scale flows [20, 21, 46]

$$\upsilon \frac{\partial^2 u}{\partial y^2} - \frac{\phi \upsilon}{k_0} u + G_x = 0, \tag{21}$$

where $u$ is the real velocity in $y$-section, $k_0$ is the medium permeability that is characterized by the dimensionless Darcy number ($Da = k_0/H^2$) ranging from $10^{-6}$ to $10^{-2}$ as $\phi$ varies from 0.1 to 0.5. Under different boundary conditions, the velocity profiles of porous Poiseuille flow (I) and porous Couette flow (II) can be analytically solved from Eq. (21) [20, 29]

$$\begin{aligned}&(\text{I}) \ G_x = 10^{-6}, \ u_{y=H} = 0, \ u_{y=H} = 0, \ u(y) = \frac{G_x k_0}{\phi \upsilon}\left(1 - \frac{\cosh\left(\sqrt{\phi/k_0}(y-H/2)\right)}{\cosh(\sqrt{\phi/k_0}H/2)}\right),\\ &(\text{II}) \ G_x = 0, \ u_{y=0} = 0, \ u_{y=H} = 10^{-3}, \ u(y) = u_{y=H}\frac{\sinh\left(\sqrt{\phi/k_0}\,y\right)}{\sinh(\sqrt{\phi/k_0}H)}.\end{aligned} \tag{22}$$

In LB coding, the no-slip boundary condition at the solid wall and the velocity condition in Couette flow are respectively implemented by the half bounce-back and the non-equilibrium extrapolation scheme [47]. As illustrated in Fig. 1 (a) and (b), it is observed that all simulation results are in accordance with their analytical solutions for different values of $Da$ or $\phi$, supporting that MRTLB-VANSE is capable of reproducing the REV-scale flow at various volume-averaging levels.

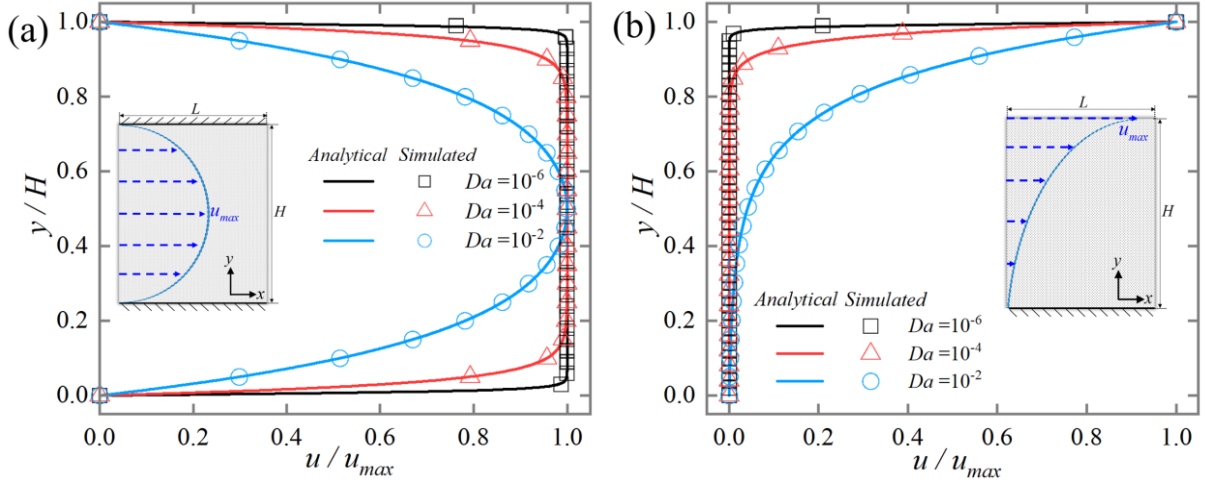

**Fig. 1.** Normalized velocity profiles and the illustrations of simulation scenarios for the porous Poiseuille flow (a) and Couette flow (b) within a uniform porosity field, with the computational domain size set as $L \times H = 100 \times 50$.

### 3.2 Nonuniform particle flow

To validate the pressure gradient term restored from MRTLB-VANSE, a simplistic scenario of 1-D nonuniform particle flow within a spatially and temporally fluctuating void fraction field is considered [29, 44]. Given the numerical error of Eq. (12) becomes nonzero in the presence of void fraction gradients, this test serves as evidence for the accuracy of the improved scheme. By artificially constructing a specific fluid-solid drag law as well as the void fraction function, the velocity distribution can be analytically solved, which are expressed as follows

$$\phi(x,t) = 1 + a\left(\sin\frac{2\pi}{L}(x - u_s t) - 1\right),$$
$$F_x^B(x,t) = \phi\left(G_x - \phi(1-\phi)(u - u_s)\right), \quad (23)$$
$$u(x,t) = u_s + \frac{G_x}{a\phi},$$

where the external force is fixed at $G_x = 10^{-6}$, $a$ and $u_s$ are constants that control the amplitude of void fraction fluctuation in space and time, respectively.
$t = 5 \times 10^6$ (b), with the computational domain set as $L \times H = 100 \times 5$.

Two verification cases are displayed in the schematic diagrams of Fig. 2 (a) and (b): one steady-state ($u_s = 0$) and one transient ($a = 0.45$). As clearly demonstrated, the simulated results align with the analytical solution of Eq. (23) under fluctuations of different amplitudes (i.e., $a = 0.15 \sim 0.45$, $u_s = 10^{-5} \sim 10^{-3}$), confirming that the pressure correction in MRTLB-VANSE is consistent and reliable.

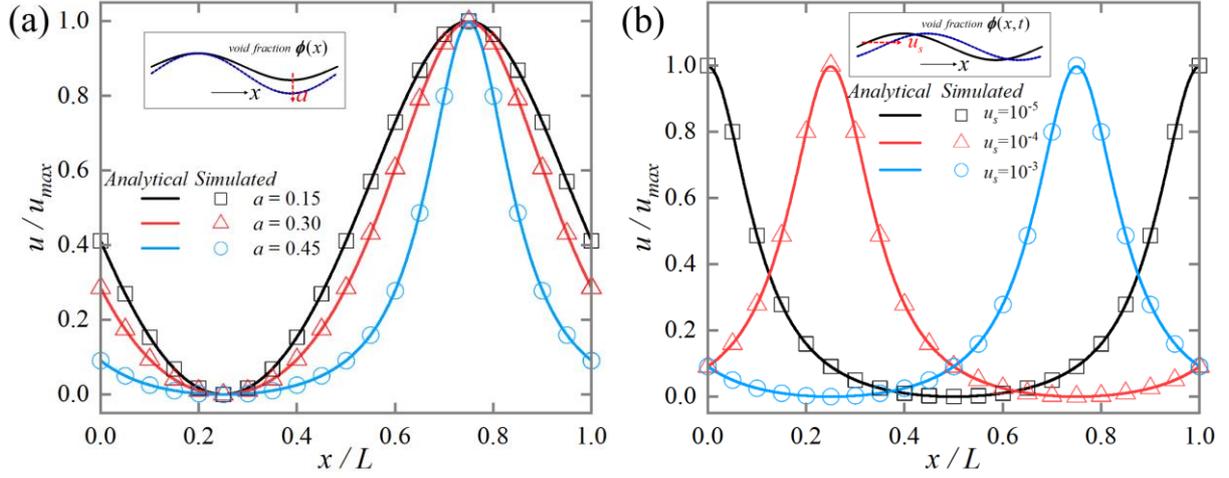

**Fig. 2.** Normalized velocity profiles of the 1-D nonuniform particle flow and schematic diagrams of the void fraction distribution for the steady-state case (a) and the transient case at

**3.3 Flow in discontinuous field**

Different from the cases in Sec. 3.2 where the void fraction is continuously distributed, this subsection conducts numerical tests under discontinuous cases. In an initially quiescent flow field, the velocity distribution theoretically remains zero under no external driving force and inlet flow. However, the previous VANSE-LB scheme suffers from significant spurious velocities in the presence of void fraction gradients, primarily caused by the numerical error of the discretized correction force, i.e., Eq. (12), which leads to non-physical acceleration on fluid. To evaluate the ability of MRTLB-VANSE to relieve this issue, a discontinuous void fraction field is configured to conduct a no-flow verification

$$\phi(x,y) = \begin{cases} \phi_0, & \forall (x,y) \in \Omega_s = [0.4L, 0.6L] \times [0.4H, 0.6H] \\ 1, & \forall (x,y) \in \Omega / \Omega_s \end{cases}, \quad (24)$$

where the 2-D computational domain $\Omega$ is set with the size $L \times H = 100 \times 100$, $\phi_0$ is a constant less than 1, thus forming a discontinuous field.

As shown in Fig. 3, the maximum spurious velocities simulated by MRTLB-VANSE for domains with different $\phi_0$ are compared with the previous density-based schemes, Zhang et al. [25], Bukreev et al. [31] and Blais et al. [26]. It is evident that although the scheme of Bukreev et al. provides some improvement and that of Blais et al. reduces the spurious velocities obviously, they cannot suit for large void fraction gradients when $\phi_0 \leq 0.3$. Exceptionally, MRTLB-VANSE demonstrates relatively low spurious velocities in the no-flow tests across different discontinuous cases. This good numerical performance can be attributed not only to the new corrected force of Eq. (13) but also, more significantly, to the redefined density equilibrium distribution of Eq. (14), which means the density distributions for $i \neq 0$ are no longer dependent on local void fraction, ensuring that the initialized $f_{i \neq 0}$ still preserve isotropy within a non-uniform void fraction field after the streaming step. Consequently, the first moment of the density distribution approximates zero, thereby producing the very low spurious velocities. Recently, Fu et al. [1] even achieved exactly-zero spurious velocities in their no-flow tests, where the influence of weak compressibility on flow field can be avoided by employing the pressure-based equilibrium distribution. Although we cannot entirely overcome the generation of spurious velocities due to the natural limitations of weak compressibility in density-based LB method, the designed MRTLB scheme has significantly improved the numerical performance of VANSE-LB model.

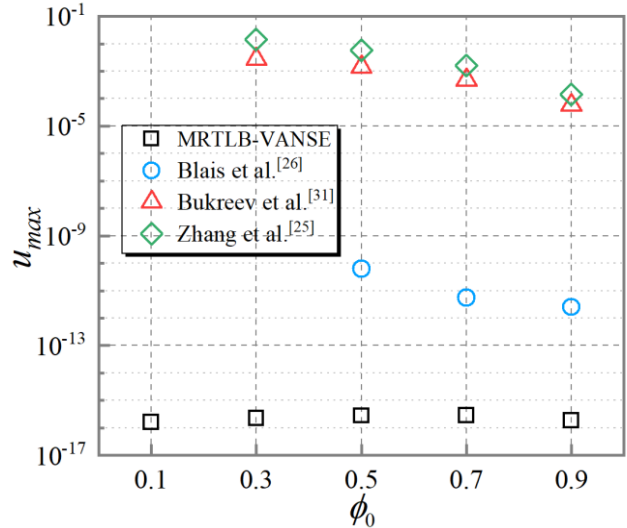

**Fig. 3.** Maximum spurious velocities of within the computational domain in the no-flow test

Furthermore, the discontinuous flow is tested by applying a external force $G_x = 10^{-6}$ towards the computational domain $\Omega$ described with Eq. (24). First, the block area of $\Omega_s$ is set as solid with $\phi_0 = 0$, and the flow field under this impermeable condition is served as a reference case as shown in Fig. 4(a). Subsequently, $\phi_0$ is adjusted to 0.5 and a high Darcy resistance ($Da = 10^{-6}$) is exerted for $\Omega_s$ to hinder fluid flow. Ideally, the flow fields obtained from the impermeable and high-resistance configurations ought to be similar. As exhibited in Fig. 4, the velocity distributions simulated by the schemes of Zhang et al. and Bukreev et al. significantly deviate from the reference case due to unacceptable spurious velocities. Although Blais et al.'s scheme provides consistent results, it fails to converge when $\phi_0 < 0.5$. The scheme of Fu et al. and MRTLB-VANSE both produce results for $\phi_0 < 0.5$, nonetheless, Fig. 4(e) shows velocity fluctuations when $\phi_0 = 0.2$ due to the presence of large pressure gradient at the free-permeable interface there. Obviously, MRTLB-VANSE performs well at $\phi_0 = 0.5, 0.2$ and yields velocity fields in accordance with the reference case, which reaffirms its numerical robustness to eliminate spurious currents in discontinuous field.

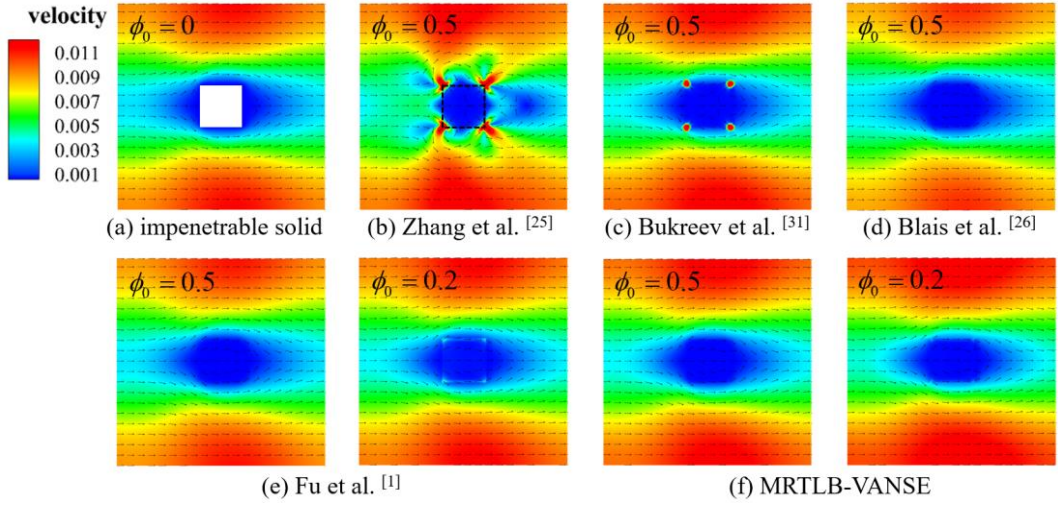

Fig. 4. The velocity distribution of forced discontinuous flow, where the domain center is arranged as a block with the constant void fraction, as indicated by the black dashed line in (a).

### 3.4 Convergence test

The method of manufactured solutions (MMS) is a straightforward approach to constructing analytical solutions for partial differential equations [34, 35], which has been widely assumed as a convincing procedure to test the convergence accuracy of a LB scheme for solving VANSE [1, 26, 31]. Specifically, the settings of velocity $\tilde{\mathbf{u}}$ and void fraction $\tilde{\phi}$ are first manufactured to meet the continuity equation in Eq. (1), and then a body force is further applied in Eq. (6) to satisfy momentum conservation

$$\mathbf{F}^B(\tilde{\mathbf{u}},\tilde{\phi}) = \partial_t(\tilde{\phi}\rho\tilde{\mathbf{u}}) + \nabla\cdot(\tilde{\phi}\rho\tilde{\mathbf{u}}\tilde{\mathbf{u}}) + \tilde{\phi}\nabla p - \upsilon\nabla\cdot\left(\tilde{\phi}\rho\left((\nabla\tilde{\mathbf{u}}) + (\nabla\tilde{\mathbf{u}})^{\mathrm{T}} - \frac{2}{3}(\nabla\cdot\tilde{\mathbf{u}})\mathbf{I}\right)\right), \qquad (25)$$

which includes all terms of the momentum equation and makes the manufactured settings become the theoretical exact solution for VANSE. To monitor the numerical results of MRTLB-VANSE, the Euclidean norm of the relative velocity is introduced

$$E_u = \sqrt{\frac{\sum_i^N |\mathbf{u}_i - \tilde{\mathbf{u}}_i|^2}{\sum_i^N |\tilde{\mathbf{u}}_i|^2}}, \qquad (26)$$

where $\mathbf{u}_i$ and $\tilde{\mathbf{u}}_i$ denote the numerical and manufactured values, respectively, $N = L \times H$ is the total number of lattice nodes. Due to the use of periodic boundary conditions towards the computational domain, the error norm is only recorded as an asymptotically stable value after sufficiently long simulation steps. The macroscopic size of the test domain in this section is fixed as $1\times 1$, with the same number of lattice nodes in $x$ and $y$ directions (i.e., $L = H = n$), and the unit length per lattice is denoted as $\Delta x = 1/n$.

Taking different cases of void fraction with spatial and spatiotemporal variations into account, three manufactured settings are designed as follows

(I)
$$\tilde{\phi} = a + b\sin(\pi x^*)\sin(\pi y^*),$$
$$\tilde{\mathbf{u}} = u_s \Delta x \begin{bmatrix} -\sin^2(\pi x^*)\sin(\pi y^*)\cos(\pi y^*) \\ \sin^2(\pi y^*)\sin(\pi x^*)\cos(\pi x^*) \end{bmatrix},$$
(27)

(II)
$$\tilde{\phi} = a\exp\left(-b\sin(\pi x^*)\sin(\pi y^*)\right),$$
$$\tilde{\mathbf{u}} = u_s \Delta x \exp\left(b\sin(\pi x^*)\sin(\pi y^*)\right)\begin{bmatrix} 1 \\ 1 \end{bmatrix},$$
(28)

(III)
$$\tilde{\phi} = a + b\sin(\pi x_t^*)\sin(\pi y_t^*),$$
$$\tilde{\mathbf{u}} = u_s \Delta x \left(\frac{1}{\tilde{\phi}} - 2\right)\begin{bmatrix} 1 \\ 1 \end{bmatrix},$$
(29)

where $a$, $b$ are constants responsible for the spatial distribution of void fraction, $u_s$ is the reference velocity, and the coordinates are given in lattice units

$$x^* = \frac{2x-n+1}{n},\ y^* = \frac{2y-n+1}{n},$$
$$x_t^* = \frac{2x-n+1}{n} + \frac{4u_s t}{n^2},\ y_t^* = \frac{2y-n+1}{n} + \frac{4u_s t}{n^2},$$
(30)

Then, substituting these manufactured continuous functions into Eq. (25), the MMS body force can be obtained through some derivative operations, as detailed in Appendix C. Note that since the density is initialized as a constant value, the pressure gradient term in the MMS body force is ignored, while the pressure correction force of Eq. (13) still takes effects in simulation.

The void fraction and steady-state velocity distributions of two stationary cases I and II are shown in Figs. 5 and 7, respectively, where $u_s$ is both fixed at 0.8. Their void fraction fields only vary in space, and the difference between them is that the velocity field of case II is non-divergence-free, resulting in all terms of viscous stress tensor in Eq. (25) are non-zero. As presented in Figs. 6 and 8, it is observed that all slopes of the fitted lines are very close to 2, i.e., the simulated velocity errors $E_u$ exhibit complete second-order convergence with the lattice spacing $\Delta x$. These results demonstrate that MRTLB-VANSE is validly without loss of accuracy for both divergence-free and non-divergence-free flows at different kinematic viscosities $\upsilon$ and large void fraction gradients.

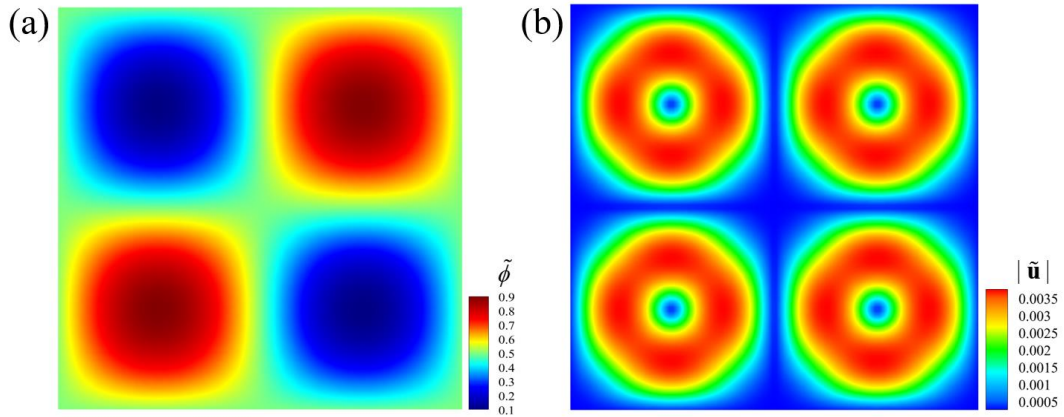

**Fig. 5.** A demo distributions of void fraction (a) and velocity (b) for case I with $a=0.5, b=0.4$, $n=100$, where the flow field is divergence-free.

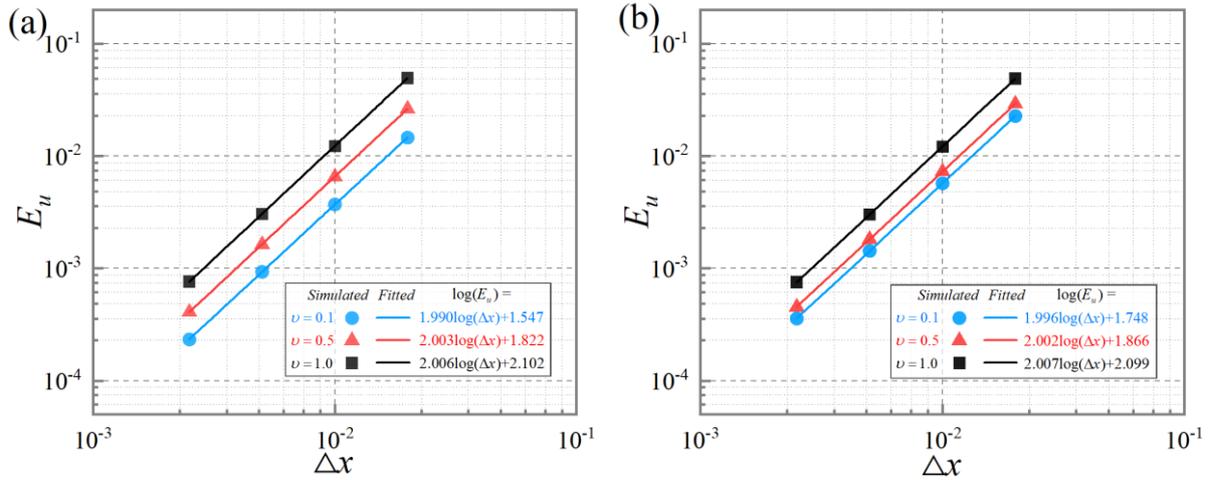

**Fig. 6.** Euclidean norm of velocity error as a function of lattice spacing for case I with two void fraction gradients: (a) $a=0.5, b=0.4$, and (b) $a=0.5, b=0.49$.

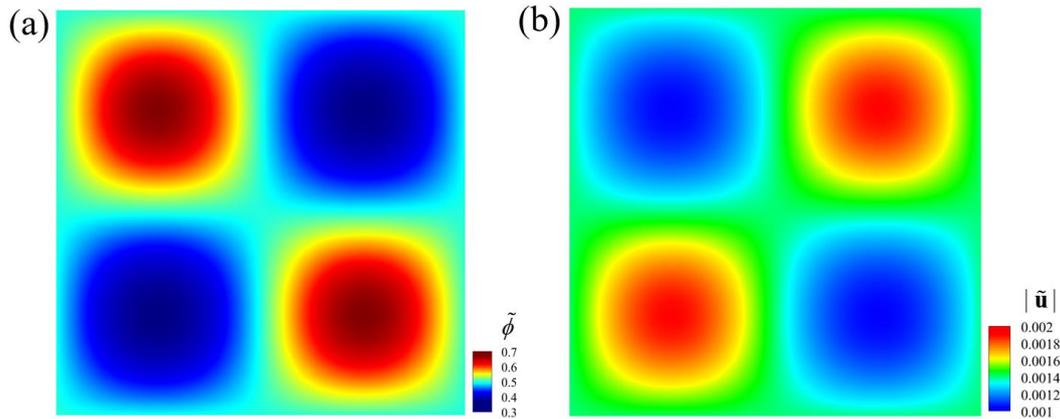

**Fig. 7.** A demo distributions of void fraction (a) and velocity (b) for case II with $a=0.5, b=0.3$, $n=100$, where the flow field is non-divergence-free.

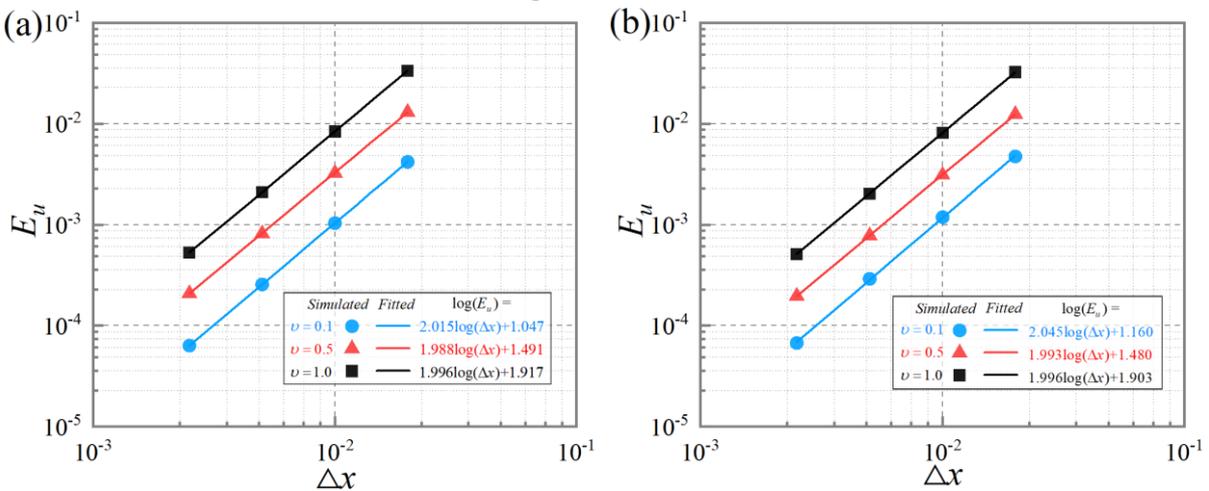

**Fig. 8.** Euclidean norm of velocity error as a function of lattice spacing for case II with two void fraction gradients: (a) $a=0.5, b=0.3$, and (b) $a=0.5, b=0.5$.

To further validate the accuracy for recovering full form of VANSE, the most complex transient case III is tested with temporal-varying void fraction field. As displayed in Fig. 8, the manufactured void fraction and velocity field is movable in *x* and *y* directions with a period of $T = n^2/|4u_s|$, in which $u_s$ is fixed at 0.08. In this case, the latest pressure-based method of Fu et al. [1] is additionally recalled for comparison. In Fig. 9, it is observed that MRTLB-VANSE exhibits almost second-order convergence under different viscosities and void fraction gradients, what's more, its error norms (solid samples) show lower absolute values than those simulated from Fu et al.'s scheme (hollow samples), relatively. Although Fu et al.'s scheme satisfies the second-order convergence at low viscosity ($\upsilon = 0.1$), it gradually loses accuracy at high viscosities ($\upsilon = 1.0$) due to the inherent instability of the SRT collision operator for relaxation times less than 0.5 [15]. This issue can be effectively addressed by the MRT collision operator, which endows MRTLB-VANSE with better numerical performance.

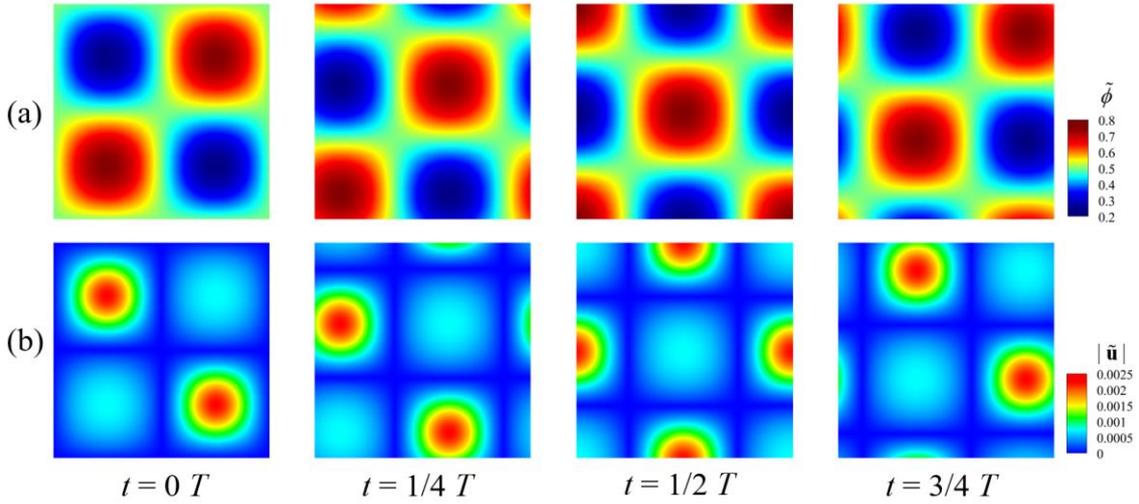

**Fig. 8.** A demo spatiotemporal distributions of void fraction (a) and velocity (b) for case III with $a=0.5, b=0.25, n=100$, where the flow field is non-divergence-free.

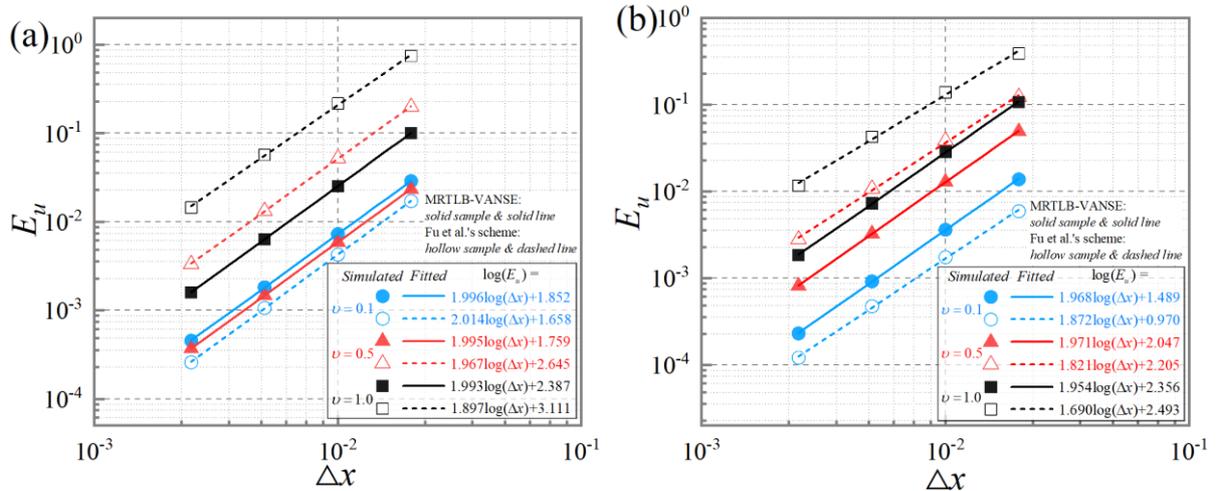

**Fig. 9.** Euclidean norm of velocity error as a function of lattice spacing for case III with two void fraction gradients: (a) $a=0.5, b=0.25$, (b) $a=0.5, b=0.45$, where the solid samples and solid lines are results of MRTLB-VANSE, and the hollow samples and dashed lines are obtained from the scheme of Fu et al.

15 / 23

## 4. Conclusions

In this paper, a lattice Boltzmann method based on the MRT collision operator is developed to asymptotically solve VANSE. By adjusting the density equilibrium distribution, a provisional equation of state is incorporated to decouple the void fraction from density. On the one hand, this decoupling not only gets rid of computing void fraction gradients in the pressure correction force, but also mitigates the force interference on flow field with the introduced numerical constant, thereby reducing non-physical oscillations caused by discretization errors at discontinuous distributions of void fraction. On the other hand, this decoupling ensures the population distributions remain isotropic when they stream in lattice space, which effectively suppresses spurious velocities calculated from their 1st-order moments. What's more, the Galilean invariance of the recovered VANSE is guaranteed by applying a penalty source term in moment space to eliminate unwanted numerical errors of the viscous stress term.

This novel method is exactly confirmed by Chapman-Enskog analysis, and detailed numerical experiments are conducted to validate its effectiveness for VANSE, including analytical benchmarks (REV-scale porous flows, 1-D particle flow), no-flow tests in discontinuous void fraction fields, as well as the convergence validation through the method of manufactured solutions. These results convincingly demonstrate that the proposed method is capable of recovering consistent VANSE with second-order accuracy, and suitable for void fraction fields with large gradients and spatiotemporal distributions. In comparison with previous methods, it shows better numerical performance, significantly advancing the applicability of the density-based VANSE-LB schemes.

In future, extending this method to 3-D is straightforward, and considering the inherent parallelizability and computational efficiency of LBM, it is promising to improve the reliability of industrial models related to VANSE. Furthermore, coupling concentration fields, temperature fields, immiscible multiphase models (color-gradient model, pseudo-potential model, phase-field model), and miscible fluid-solid models (TFM, DEM), numerous mesoscopic methods with engineering application potential can be developed to simulate complex liquid-gas-solid processes.

**Acknowledgements**

This research is funded by the National Key R&D Program of China (No. 2019YFA0708704).16 / 23

**Appendix A. Chapman-Enskog analysis**

In this appendix, the Chapman-Enskog expansion is derived in detail to demonstrate that the established MRT-LB scheme can recover the consistent VANSE.

First, the 9×9 orthogonal matrix and its inverse matrix are presented

$$\mathbf{M} = \begin{pmatrix} 1 & 1 & 1 & 1 & 1 & 1 & 1 & 1 & 1 \\ -4 & -1 & -1 & -1 & -1 & 2 & 2 & 2 & 2 \\ 4 & -2 & -2 & -2 & -2 & 1 & 1 & 1 & 1 \\ 0 & 1 & 0 & -1 & 0 & 1 & -1 & -1 & 1 \\ 0 & -2 & 0 & 2 & 0 & 1 & -1 & -1 & 1 \\ 0 & 0 & 1 & 0 & -1 & 1 & 1 & -1 & -1 \\ 0 & 0 & -2 & 0 & 2 & 1 & 1 & -1 & -1 \\ 0 & 1 & -1 & 1 & -1 & 0 & 0 & 0 & 0 \\ 0 & 0 & 0 & 0 & 0 & 1 & -1 & 1 & -1 \end{pmatrix}, \quad \mathbf{M}^{-1} = \frac{1}{36}\begin{pmatrix} 4 & -4 & 4 & 0 & 0 & 0 & 0 & 0 & 0 \\ 4 & -1 & -2 & 6 & -6 & 0 & 0 & 9 & 0 \\ 4 & -1 & -2 & 0 & 0 & 6 & -6 & -9 & 0 \\ 4 & -1 & -2 & -6 & 6 & 0 & 0 & 9 & 0 \\ 4 & -1 & -2 & 0 & 0 & -6 & 6 & -9 & 0 \\ 4 & 2 & 1 & 6 & 3 & 6 & 3 & 0 & 9 \\ 4 & 2 & 1 & -6 & -3 & 6 & 3 & 0 & -9 \\ 4 & 2 & 1 & -6 & -3 & -6 & -3 & 0 & 9 \\ 4 & 2 & 1 & 6 & 3 & -6 & -3 & 0 & -9 \end{pmatrix}. \quad \text{(A1)}$$

Then, the evolution of MRT-LB equation in velocity space and moment space are respectively expressed as follows

$$f_i(\mathbf{x}+\mathbf{e}_i\delta_t, t+\delta_t) = f_i(\mathbf{x},t) + \sum_j (\mathbf{M}^{-1}\mathbf{\Gamma})_{ij}(m_j - m_j^{eq}) + \sum_j \left(\mathbf{M}^{-1}(\mathbf{I}-\frac{\mathbf{\Gamma}}{2})\right)_{ij}(C_j + S_j), \quad \text{(A2)}$$

$$\mathbf{m}(\mathbf{x}+\mathbf{e}_i\delta_t, t+\delta_t) = \mathbf{m}(\mathbf{x},t) - \mathbf{\Gamma}\left(\mathbf{m}(\mathbf{x},t) - \mathbf{m}^{eq}(\mathbf{x},t)\right) + (\mathbf{I}-\frac{\mathbf{\Gamma}}{2})(\mathbf{C}(\mathbf{x},t)+\mathbf{S}(\mathbf{x},t)). \quad \text{(A3)}$$

The moment function, the derivatives of time and space, as well as the source term and force term are expanded at consecutive scales of $\epsilon$

$$\mathbf{m} = \mathbf{m}^{(0)} + \epsilon\mathbf{m}^{(1)} + \epsilon^2\mathbf{m}^{(2)} + \cdots, \quad \partial_t = \epsilon\partial_{t1} + \epsilon^2\partial_{t2}, \quad \partial_\alpha = \epsilon\partial_{\alpha 1}, \quad \mathbf{C} = \epsilon\mathbf{C}^1, \quad \mathbf{S} = \epsilon\mathbf{S}^1, \quad \text{(A4)}$$

where the external body force $\mathbf{F}^B$ is ignored in $\mathbf{S}$ for brevity; when $n > 0$, $m_0^{(n)} = (\phi\rho)^{(n)} = 0$, $m_4^{(n)} = j_x^{(n)} = -\delta_t F_x^p/2$, $m_6^{(n)} = j_y^{(n)} = -\delta_t F_y^p/2$ are obtained since $\phi\rho$ and $\phi\rho\mathbf{u}$ are conservative variables. Substituting the above expressions into Eq. (A3) and applying Taylor expansion, we can obtain the zero-, first-, and second-order equations in $\epsilon$

$$\epsilon^0: \mathbf{m}^{(0)} = \mathbf{m}^{eq},$$

$$\epsilon^1: \tilde{\mathbf{D}}_1\mathbf{m}^{(0)} = -\mathbf{\Gamma}'\mathbf{m}^{(1)} + \left(\mathbf{I}-\frac{\mathbf{\Gamma}}{2}\right)\mathbf{M}(\mathbf{C}^1+\mathbf{S}^1), \quad \text{(A5)}$$

$$\epsilon^2: \partial_{t2}\mathbf{m}^{(0)} + \tilde{\mathbf{D}}_1\left(\mathbf{I}-\frac{\mathbf{\Gamma}}{2}\right)\mathbf{m}^{(1)} + \frac{\delta_t}{2}\tilde{\mathbf{D}}_1\left(\mathbf{I}-\frac{\mathbf{\Gamma}}{2}\right)\mathbf{M}(\mathbf{C}^1+\mathbf{S}^1) = -\mathbf{\Gamma}'\mathbf{m}^{(2)},$$

where $\mathbf{\Gamma}' = \mathbf{\Gamma}/\delta_t$, $\tilde{\mathbf{D}}_1 = \mathbf{M}\mathbf{D}_1\mathbf{M}^{-1} = \partial_{t1}\mathbf{I} + \mathbf{E}_\alpha\partial_{\alpha 1}$, $\mathbf{D}_1 = \partial_{t1}\mathbf{I} + \partial_{\alpha 1}\text{diag}(e_{0\alpha}, e_{1\alpha}, \cdots, e_{8\alpha})$, and $\mathbf{E}_\alpha$ are explicitly provided in literature [44].

Based on above preparations, Eq. (A5) in $\epsilon^1$ can be rewritten at $t_1$ time scale as

$$\partial_{t1}\begin{pmatrix}\phi\rho\\ \phi\rho(-4+3u_x^2+3u_y^2+2X)\\ \phi\rho(4-3u_x^2-3u_y^2-3X)\\ \phi\rho u_x\\ (X-2)\phi\rho u_x\\ \phi\rho u_y\\ (X-2)\phi\rho u_y\\ \phi\rho(u_x^2-u_y^2)\\ \phi\rho u_x u_y\end{pmatrix}+\partial_{x1}\begin{pmatrix}\phi\rho u_x\\ (X-1)\phi\rho u_x\\ (X-2)\phi\rho u_x\\ \kappa\rho c_s^2+\phi\rho u_x^2\\ -\kappa\rho c_s^2-\phi\rho(u_x^2-u_y^2)\\ \phi\rho u_x u_y\\ \phi\rho u_x u_y\\ -\kappa\rho c_s^2 u_x+\phi\rho u_x\\ \kappa\rho c_s^2 u_y\end{pmatrix}+\partial_{y1}\begin{pmatrix}\phi\rho u_y\\ (X-1)\phi\rho u_y\\ (X-2)\phi\rho u_y\\ \phi\rho u_x u_y\\ \phi\rho u_x u_y\\ \kappa\rho c_s^2+\phi\rho u_y^2\\ -\kappa\rho c_s^2+\phi\rho(u_x^2-u_y^2)\\ \kappa\rho c_s^2 u_y-\phi\rho u_y\\ \kappa\rho c_s^2 u_x\end{pmatrix}$$

$$=\begin{pmatrix}0\\ -\tau_e' e^{(1)}\\ -\tau_\varepsilon' \varepsilon^{(1)}\\ \tau_c' \delta_t F_{x1}^p/2\\ -\tau_q' q_x^{(1)}\\ \tau_c' \delta_t F_{y1}^p/2\\ -\tau_q' q_y^{(1)}\\ -\tau_\upsilon' p_{xx}^{(1)}\\ -\tau_\upsilon' p_{xy}^{(1)}\end{pmatrix}+\begin{pmatrix}0\\ 6(1-\tau_e/2)(u_x F_{x1}^p+u_y F_{y1}^p)\\ -6(1-\tau_\varepsilon/2)(u_x F_{x1}^p+u_y F_{y1}^p)\\ (1-\tau_c/2)F_{x1}^p\\ -(1-\tau_q/2)F_{x1}^p\\ (1-\tau_c/2)F_{y1}^p\\ -(1-\tau_q/2)F_{y1}^p\\ 2(1-\tau_\upsilon/2)(u_x F_{x1}^p-u_y F_{y1}^p)\\ (1-\tau_\upsilon/2)(u_x F_{y1}^p+u_y F_{x1}^p)\end{pmatrix}+\begin{pmatrix}0\\ (1-\tau_e/2)C_{a1}\\ 0\\ 0\\ 0\\ 0\\ 0\\ (1-\tau_\upsilon/2)C_{b1}\\ (1-\tau_\upsilon/2)C_{c1}\end{pmatrix},\quad(A6)$$

Similarly, Eq. (A5) in $\epsilon^2$ can also be derived at $t_2$ time scale and the components corresponding to the conservative moments are given by

$$\partial_{t2}(\phi\rho)=0,$$

$$\partial_{t2}(\phi\rho u_x)-\frac{\delta_t}{2}\partial_{t1}\left[(1-\frac{\tau_c}{2})F_{x1}^p\right]+\frac{1}{6}\partial_{x1}\left[(1-\frac{\tau_e}{2})e^{(1)}\right]+\frac{1}{2}\partial_{x1}\left[(1-\frac{\tau_\upsilon}{2})p_{xx}^{(1)}\right]+\partial_{y1}\left[(1-\frac{\tau_\upsilon}{2})p_{xy}^{(1)}\right]$$

$$+\frac{\delta_t}{2}\left\{\partial_{t1}\left[(1-\frac{\tau_c}{2})F_{x1}^p\right]+\partial_{x1}\left[\frac{\tau_\upsilon}{2}(u_y F_{y1}^p-u_x F_{x1}^p)-\frac{\tau_e}{2}(u_x F_{x1}^p+u_y F_{y1}^p)+2u_x F_{x1}^p\right]+\partial_{y1}\left[(1-\frac{\tau_\upsilon}{2})(u_x F_{y1}^p+u_y F_{x1}^p)\right]\right\}$$

$$+\frac{\delta_t}{2}\left\{\frac{1}{6}\partial_{x1}\left[(1-\frac{\tau_e}{2})C_{a1}\right]+\frac{1}{2}\partial_{x1}\left[(1-\frac{\tau_\upsilon}{2})C_{b1}\right]+\partial_{y1}\left[(1-\frac{\tau_\upsilon}{2})C_{c1}\right]\right\}=0,$$

$$\partial_{t2}(\phi\rho u_y)-\frac{\delta_t}{2}\partial_{t1}\left[(1-\frac{\tau_c}{2})F_{y1}^p\right]+\frac{1}{6}\partial_{y1}\left[(1-\frac{\tau_e}{2})e^{(1)}\right]-\frac{1}{2}\partial_{y1}\left[(1-\frac{\tau_\upsilon}{2})p_{xx}^{(1)}\right]+\partial_{x1}\left[(1-\frac{\tau_\upsilon}{2})p_{xy}^{(1)}\right]$$

$$+\frac{\delta_t}{2}\left\{\partial_{t1}\left[(1-\frac{\tau_c}{2})F_{y1}^p\right]+\partial_{y1}\left[\frac{\tau_\upsilon}{2}(u_x F_{x1}^p-u_y F_{y1}^p)-\frac{\tau_e}{2}(u_x F_{x1}^p+u_y F_{y1}^p)+2u_y F_{y1}^p\right]+\partial_{x1}\left[(1-\frac{\tau_\upsilon}{2})(u_x F_{y1}^p+u_y F_{x1}^p)\right]\right\}$$

$$+\frac{\delta_t}{2}\left\{\frac{1}{6}\partial_{y1}\left[(1-\frac{\tau_e}{2})C_{a1}\right]-\frac{1}{2}\partial_{y1}\left[(1-\frac{\tau_\upsilon}{2})C_{b1}\right]+\partial_{x1}\left[(1-\frac{\tau_\upsilon}{2})C_{c1}\right]\right\}=0.$$

(A7)

Combining the correction force of Eq. (13) with the first, fourth, and sixth components of Eq. (A6), the macroscopic governing equations at $t_1$ time scale can be derived

$$\partial_{t1}(\phi\rho)+\partial_{x1}(\phi\rho u_x)+\partial_{y1}(\phi\rho u_y)=0,$$
$$\partial_{t1}(\phi\rho u_x)+\partial_{x1}(\phi\rho u_x u_x)+\partial_{y1}(\phi\rho u_x u_y)=-\phi\partial_{x1}(\rho c_s^2),\quad(A8)$$
$$\partial_{t1}(\phi\rho u_y)+\partial_{x1}(\phi\rho u_x u_y)+\partial_{y1}(\phi\rho u_y u_y)=-\phi\partial_{y1}(\rho c_s^2).$$

Subsequently, neglecting the term of order O($|\mathbf{u}|^3$) and higher-order terms of $u_\beta \partial_\alpha (u_\alpha u_\beta)$ under the incompressible condition, the expressions of $e^{(1)}$, $p_{xx}^{(1)}$ and $p_{xy}^{(1)}$ in Eq. (A7) can be obtained from the second, eighth, and ninth components of Eq. (A6)

$$-\tau'_e e^{(1)} = -2\kappa(u_x \partial_{x1}\rho + u_y \partial_{y1}\rho) + \kappa[\partial_{x1}(\rho u_x) + \partial_{y1}(\rho u_y)]$$
$$+ 3[\partial_{x1}(\phi \rho u_x) + \partial_{y1}(\phi \rho u_y)] + 2\kappa \partial_{t1}\rho + 3\tau_e(u_x F_{x1}^p + u_y F_{y1}^p) + (\frac{\tau_e}{2} - 1)C_{a1},$$

$$-\tau'_\upsilon p_{xx}^{(1)} = -\frac{2\kappa}{3}(u_x \partial_{x1}\rho - u_y \partial_{y1}\rho) - \frac{\kappa}{3}[\partial_{x1}(\rho u_x) - \partial_{y1}(\rho u_y)] \quad (A9)$$
$$+ [\partial_{x1}(\phi \rho u_x) - \partial_{y1}(\phi \rho u_y)] + \tau_\upsilon(u_x F_{x1}^p - u_y F_{y1}^p) + (\frac{\tau_\upsilon}{2} - 1)C_{b1},$$

$$-\tau'_\upsilon p_{xy}^{(1)} = \frac{\kappa}{3}\rho(\partial_{x1}u_y + \partial_{y1}u_x) + \frac{\tau_\upsilon}{2}(u_x F_{y1}^p + u_y F_{x1}^p) + (\frac{\tau_\upsilon}{2} - 1)C_{c1},$$

where no expression is provided for the time derivative of $\rho$, so we replace it with the known partial derivatives, i.e., $\partial_{t1}\rho = (\partial_{t1}(\phi\rho) - \rho\partial_{t1}\phi)/\phi = -(\partial_{x1}(\phi \rho u_x) + \partial_{y1}(\phi \rho u_y) + \rho\partial_{t1}\phi)/\phi$. Then, substituting the penalty terms of Eq. (20) into Eq. (A9), $e^{(1)}$, $p_{xx}^{(1)}$ and $p_{xy}^{(1)}$ related to the viscosity stress tensor are successfully corrected

$$-\tau'_e e^{(1)} = 2\phi\rho(\partial_{x1}u_x + \partial_{y1}u_y) + 3\tau_e(u_x F_{x1}^p + u_y F_{y1}^p) + \frac{\tau_e}{2}C_{a1},$$
$$-\tau'_\upsilon p_{xx}^{(1)} = \frac{2}{3}\phi\rho(\partial_{x1}u_x - \partial_{y1}u_y) + \tau_\upsilon(u_x F_{x1}^p - u_y F_{y1}^p) + \frac{\tau_\upsilon}{2}C_{b1}, \quad (A10)$$
$$-\tau'_\upsilon p_{xy}^{(1)} = \frac{1}{3}\phi\rho(\partial_{x1}u_y + \partial_{y1}u_x) + \frac{\tau_\upsilon}{2}(u_x F_{y1}^p + u_y F_{x1}^p) + \frac{\tau_\upsilon}{2}C_{c1}.$$

Then, substituting Eq. (A10) into Eq. (A7), the macroscopic governing equations at $t_2$ time scale can be derived

$$\partial_{t2}(\phi\rho) = 0,$$
$$\partial_{t2}(\phi\rho u_x) = \partial_{x1}\left[\phi\rho\upsilon(\partial_{x1}u_x - \partial_{y1}u_y)\right] + \partial_{y1}\left[\phi\rho\upsilon(\partial_{x1}u_y + \partial_{y1}u_x)\right] + \partial_{x1}\left[\phi\rho\upsilon_e(\partial_{x1}u_x + \partial_{y1}u_y)\right], \quad (A11)$$
$$\partial_{t2}(\phi\rho u_y) = \partial_{x1}\left[\phi\rho\upsilon(\partial_{x1}u_y + \partial_{y1}u_x)\right] + \partial_{y1}\left[\phi\rho\upsilon(\partial_{y1}u_y - \partial_{x1}u_x)\right] + \partial_{y1}\left[\phi\rho\upsilon_e(\partial_{x1}u_x + \partial_{y1}u_y)\right],$$

where $\upsilon = c_s^2(\tau_\upsilon^{-1} - 0.5)\delta_t$, $\upsilon_e = c_s^2(\tau_e^{-1} - 0.5)\delta_t$ are the kinematic shear and bulk viscosities.

Lastly, combining the equations at $t_1$ and $t_2$ time scales together, i.e., Eqs. (A8) and (A11), we recover the macroscopic equations as follows

$$\partial_t(\phi\rho) + \nabla\cdot(\phi\rho\mathbf{u}) = 0,$$
$$\partial_t(\phi\rho\mathbf{u}) + \nabla\cdot(\phi\rho\mathbf{uu}) = -\phi\nabla p + \nabla\cdot(\phi\tau), \quad (A12)$$

where $p = \rho c_s^2$ is the pressure term, $\tau = \rho\upsilon(\nabla\mathbf{u} + (\nabla\mathbf{u})^T) + \rho(\upsilon_e - \upsilon)(\nabla\cdot\mathbf{u})\mathbf{I}$ is the viscous stress tensor, and the target VANSE of Eq. (1) is exactly recovered when $\upsilon_e = \upsilon/3$ is chosen.

## Appendix B. MRTLB-VANSE with 2nd-order truncated equilibrium distribution

The second-order density equilibrium function from the Hermite expansion

$$f_i^{eq}(\mathbf{x},t) = \begin{cases} \phi\rho - \sum_{i\neq 0}\omega_i\kappa\rho - \omega_0\phi\rho\dfrac{\mathbf{u}\cdot\mathbf{u}}{2c_s^2}, & i=0 \\ \omega_i\left(\kappa\rho + \phi\rho\left(\dfrac{\mathbf{e}_i\cdot\mathbf{u}}{c_s^2} + \dfrac{(\mathbf{e}_i\cdot\mathbf{u})^2}{2c_s^4} - \dfrac{\mathbf{u}\cdot\mathbf{u}}{2c_s^2}\right)\right), & i=1\sim 8 \end{cases} \quad (B1)$$

The equilibrium moment

$$\mathbf{m}^{eq} = \mathbf{M}\mathbf{f}^{eq} = (m_0^{eq}, m_1^{eq}, \cdots, m_8^{eq})^T = $$
$$\phi\rho\left(1, -4+3\mathbf{u}^2+2X, 4-3\mathbf{u}^2-3X, u_x, -u_x, u_y, -u_y, u_x^2-u_y^2, u_xu_y\right)^T. \quad (B2)$$

The penalty source term for viscous stress

$$C_a = -2\kappa[\partial_{x1}(\rho u_x) + \partial_{y1}(\rho u_y)] + 2(\kappa-\phi)\rho(\partial_{x1}u_x + \partial_{y1}u_y)$$
$$+ (4-2X)[\partial_{x1}(\phi\rho u_x) + \partial_{y1}(\phi\rho u_y)] - 2X\rho\partial_{t1}\phi,$$

$$C_b = -\dfrac{2}{3}\kappa[\partial_{x1}(\rho u_x) - \partial_{y1}(\rho u_y)] + \dfrac{2}{3}(\kappa-\phi)\rho(\partial_{x1}u_x - \partial_{y1}u_y)$$
$$+ \dfrac{2}{3}[\partial_{x1}(\phi\rho u_x) - \partial_{y1}(\phi\rho u_y)], \quad (B3)$$

$$C_c = -\dfrac{\kappa}{3}[\partial_{x1}(\rho u_y) + \partial_{y1}(\rho u_x)] + \dfrac{1}{3}(\kappa-\phi)\rho(\partial_{x1}u_y + \partial_{y1}u_x)$$
$$+ \dfrac{1}{3}[\partial_{x1}(\phi\rho u_y) + \partial_{y1}(\phi\rho u_x)].$$

## Appendix C. MMS body forces

Case I:

$$F_x^B = \dfrac{2\pi u_s^2\tilde{\phi}}{n^3}\sin^3(\pi x^*)\cos(\pi x^*)\sin^2(\pi y^*) + \dfrac{4\upsilon\pi^2 u_s b}{n^3}\sin^3(\pi x^*)\cos^3(\pi y^*)$$
$$+ \dfrac{4\upsilon\pi^2 u_s}{n^3}\sin(\pi y^*)\cos(\pi y^*)\begin{pmatrix} -6\tilde{\phi}\sin^2(\pi x^*) + 2a\cos^2(\pi x^*) \\ +5b\sin(\pi x^*)\cos^2(\pi x^*)\sin(\pi y^*) \end{pmatrix},$$

$$F_y^B = \dfrac{2\pi u_s^2\tilde{\phi}}{n^3}\sin^2(\pi x^*)\sin^3(\pi y^*)\cos(\pi y^*) \quad (C1)$$
$$- \dfrac{4\upsilon\pi^2 u_s}{n^3}\sin(\pi x^*)\cos(\pi x^*)\cos^2(\pi y^*)\left(2a + 5b\sin(\pi x^*)\sin(\pi y^*)\right)$$
$$+ \dfrac{4\upsilon\pi^2 u_s}{n^3}\cos(\pi x^*)\sin^2(\pi y^*)\begin{pmatrix} 6a\sin(\pi x^*) + 6b\sin^2(\pi x^*)\sin(\pi y^*) \\ -b\cos^2(\pi x^*)\sin(\pi y^*) \end{pmatrix}.$$

Case II:

$$F_x^B = F_y^B = \dfrac{2\pi u_s^2 ab}{n^3}\exp(b\sin(\pi x^*)\sin(\pi y^*))\sin(\pi(x^*+y^*))$$
$$- \dfrac{4\upsilon\pi^2 u_s ab}{3n^3}\left(\cos(\pi x^*)\cos(\pi y^*) - 7\sin(\pi x^*)\sin(\pi y^*)\right). \quad (C2)$$

Case III:

$$F_x^B = -\frac{8\pi u_s^2 b}{n^3}\sin\left(\pi(x_t^* + y_t^*)\right) + \frac{2\pi u_s^2 b}{n^3 \tilde{\phi}^2}\sin\left(\pi(x_t^* + y_t^*)\right)(4\tilde{\phi}^2 - 1)$$

$$+\frac{2\upsilon\pi^2 u_s b}{3n^3 \tilde{\phi}^2}\begin{pmatrix} 3b\cos(2\pi x_t^*) + 4b\cos(2\pi y_t^*) - 7b \\ +8a\cos\left(\pi(x_t^* + y_t^*)\right) - 6a\cos\left(\pi(x_t^* - y_t^*)\right) \end{pmatrix},$$

$$F_y^B = -\frac{8\pi u_s^2 b}{n^3}\sin\left(\pi(x_t^* + y_t^*)\right) + \frac{2\pi u_s^2 b}{n^3 \tilde{\phi}^2}\sin\left(\pi(x_t^* + y_t^*)\right)(4\tilde{\phi}^2 - 1)$$

$$+\frac{2\upsilon\pi^2 u_s b}{3n^3 \tilde{\phi}^2}\begin{pmatrix} 4b\cos(2\pi x_t^*) + 3b\cos(2\pi y_t^*) - 7b \\ +8a\cos\left(\pi(x_t^* + y_t^*)\right) - 6a\cos\left(\pi(x_t^* - y_t^*)\right) \end{pmatrix}.$$

(C3)